\documentstyle[11pt,newpasp,twoside,epsf]{article}
\def\edcomment#1{\iffalse\marginpar{\raggedright\sl#1\/}\else\relax\fi}
\reversemarginpar

\begin{document}
\title{A survey of Probable Open Cluster Remnants in the Galactic Disk}
\author{Sandro Villanova, Giovanni Carraro}
\affil{Department of Astronomy, Padova University, Vicolo Osservatorio 2,
I-35122, Padova, Italy}
\author{Ra\'ul de la Fuente Marcos}
\affil{Department of Physics, Suffolk University Madrid Campus,
C/ Vi\~na 3, E-28003,Madrid, Spain}

\begin{abstract}
We present preliminary results of an ongoing survey to probe the 
physical nature of a sample of open clusters (
POCRs, Bica et al. 2001)
believed to be in an advanced stage of dynamical evolution. This kind
of object is very important to constrain dynamical $N$-body models
of clusters evolution and dissolution, and to understand the formation
of the Galactic field stars population. Moreover they might represent
what remains of much larger old clusters, which are of paramount
importance to probe the early stages of the Galactic disk evolution.
In this paper we present our results for NGC~1901.
\end{abstract}

\section{NGC 1901}
NGC~1901 is a loose aggregate of stars (see Fig.~1, left panel) 
in Doradus, and it is the subject of a recent paper by Pavani
et al. (2001). They present new CCD photometry, which they combine
with proper motions and conclude that NGC~1901 is a physical aggregate,
whose features are not typical of a classical open cluster,
but better describe an open cluster remnant.
This is quite clear from the vector point diagram in Fig~1
(right panel) which shows a clear concentration of stars
having the same tangential motion.\\ 
In this study we present a radial velocity survey of the 16 brightest
stars in NGC~1901, which allow us to better constrain the cluster
membership, an derive updated estimates of its fundamental
parameters.

\begin{figure}  
\plottwo{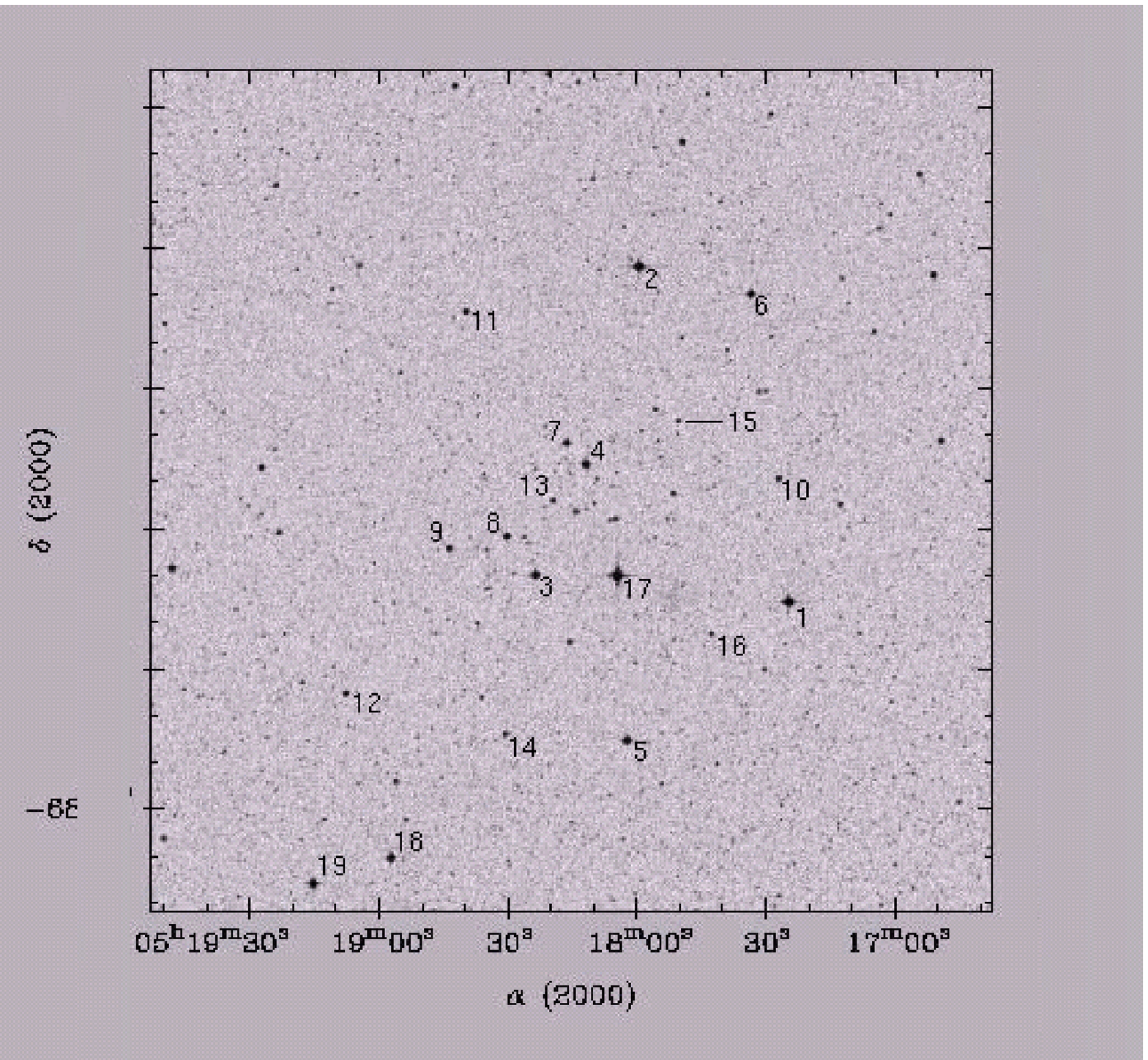}{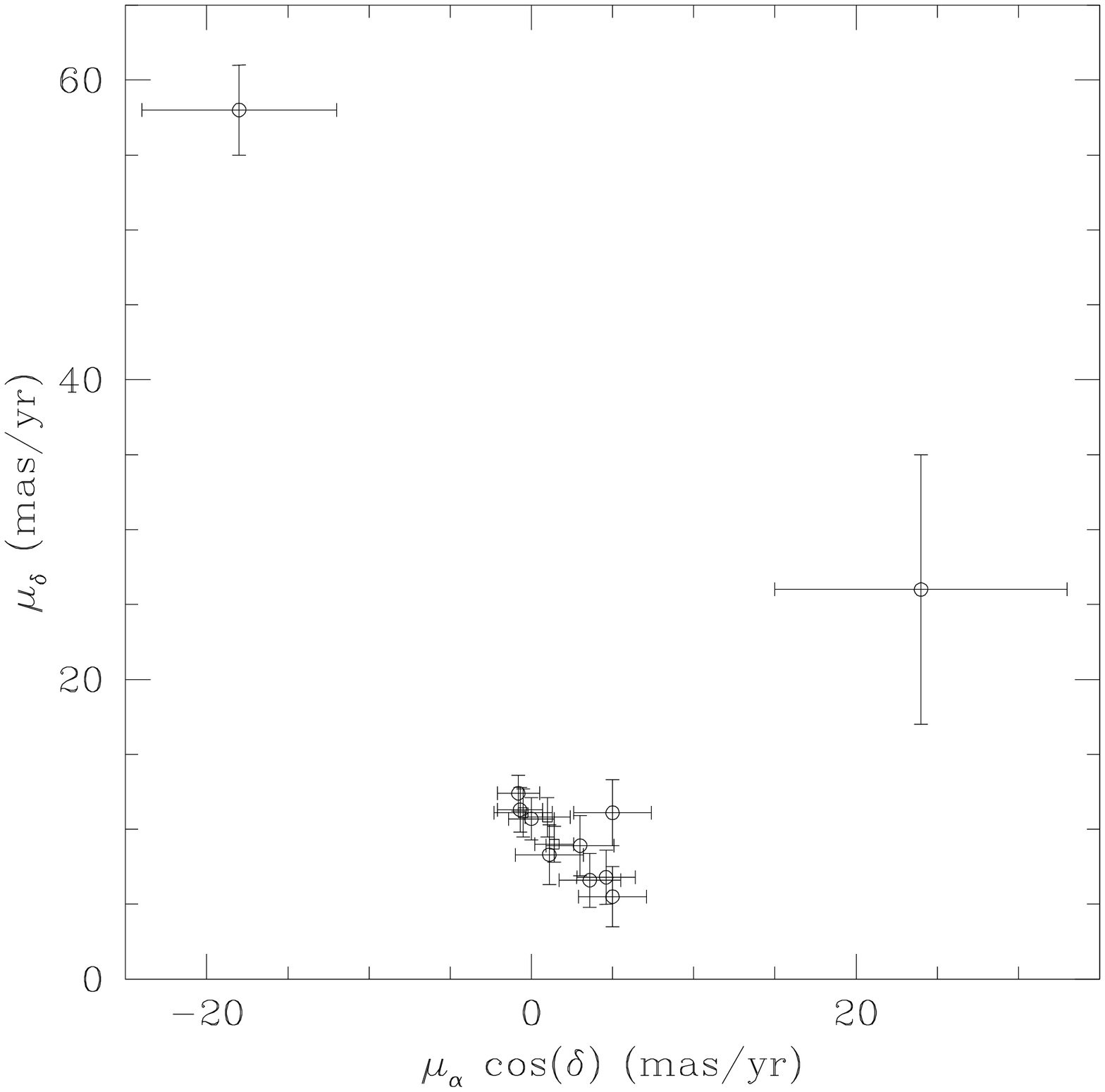}
\caption{{\bf Left}: a DSS map of NGC~1901 field.
{\bf Right}: Vector point diagram for the stars
in the field of NGC~1901.}
\end{figure}

\begin{figure}  
\plottwo{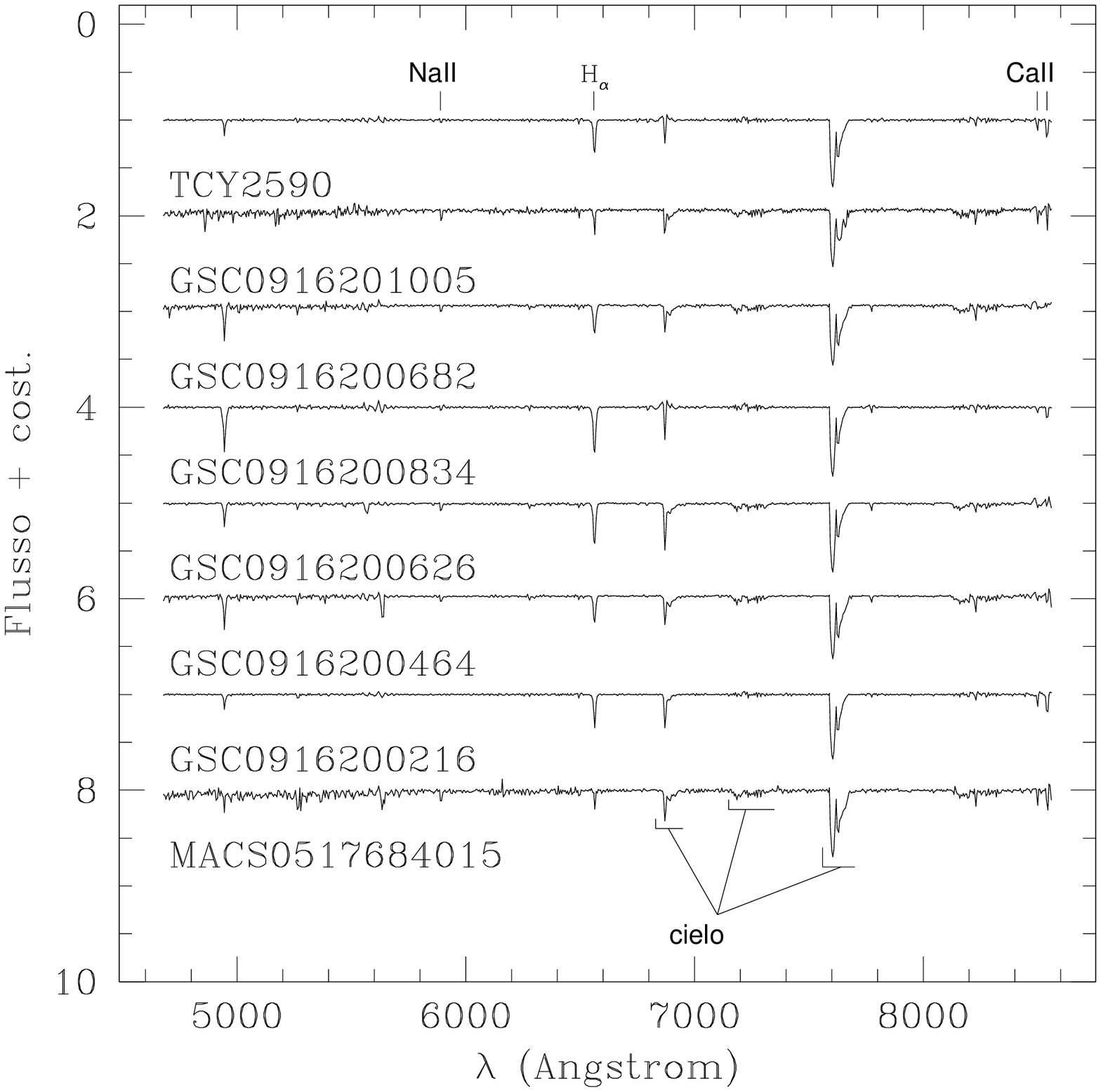}{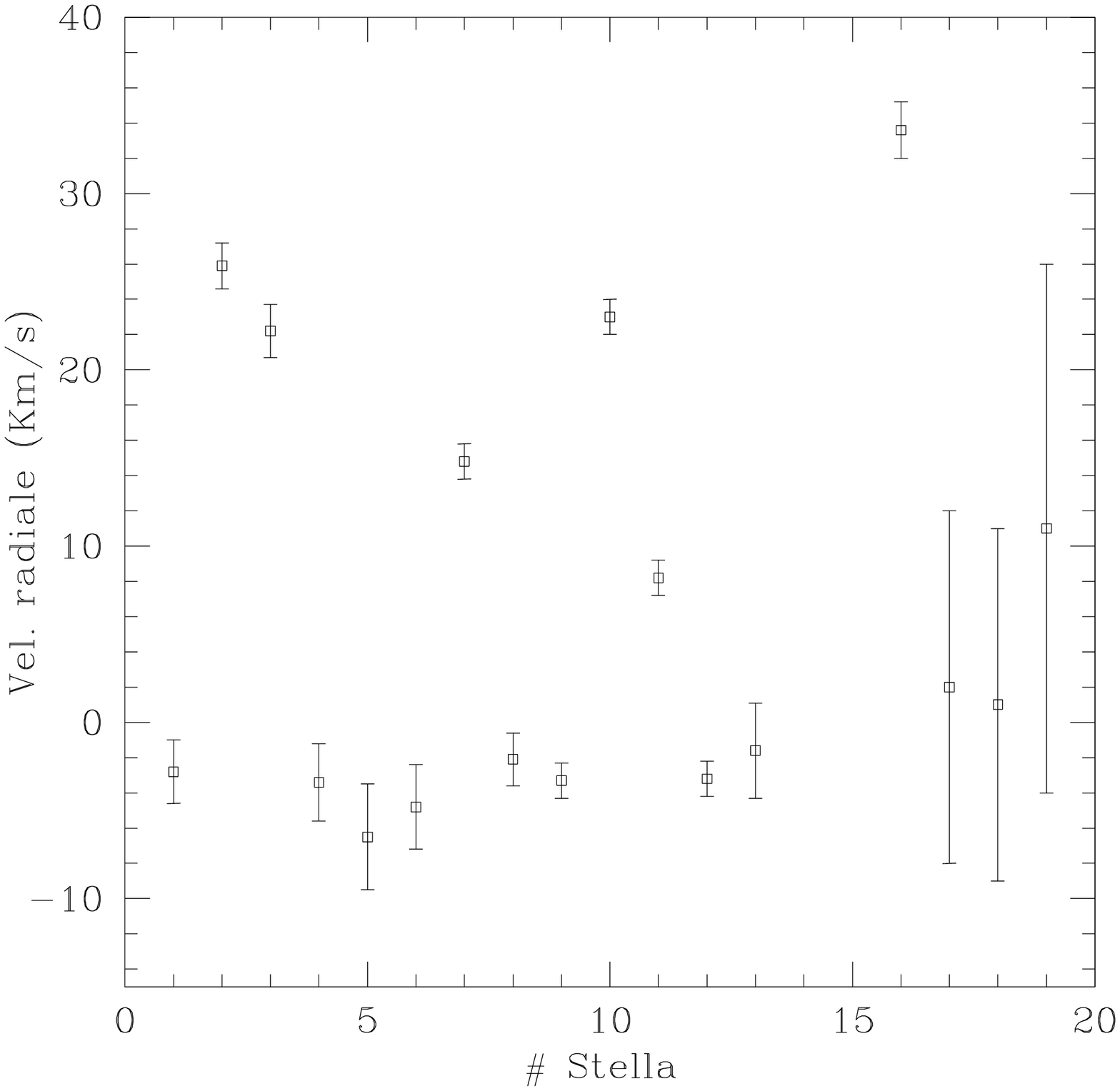}
\caption{{\bf Left}: a sample of Echelle spectra.
{\bf Right}: Radial velocity distribution of the
stars in NGC~1901.}
\end{figure}

\noindent
In Fig.~2 (left panel) we present a few example of high
resolution spectra, whereas in Fig.~2 (right panel) 
we show the radial velocity distribution which allows
- together with the vector point diagram -
to discriminate between members and non members.
We find that $v_r$ = -3.5$\pm$1.0 km/sec, and a radial
velocity dispersion of 2.4 km/sec.\\

\begin{figure}  
\plottwo{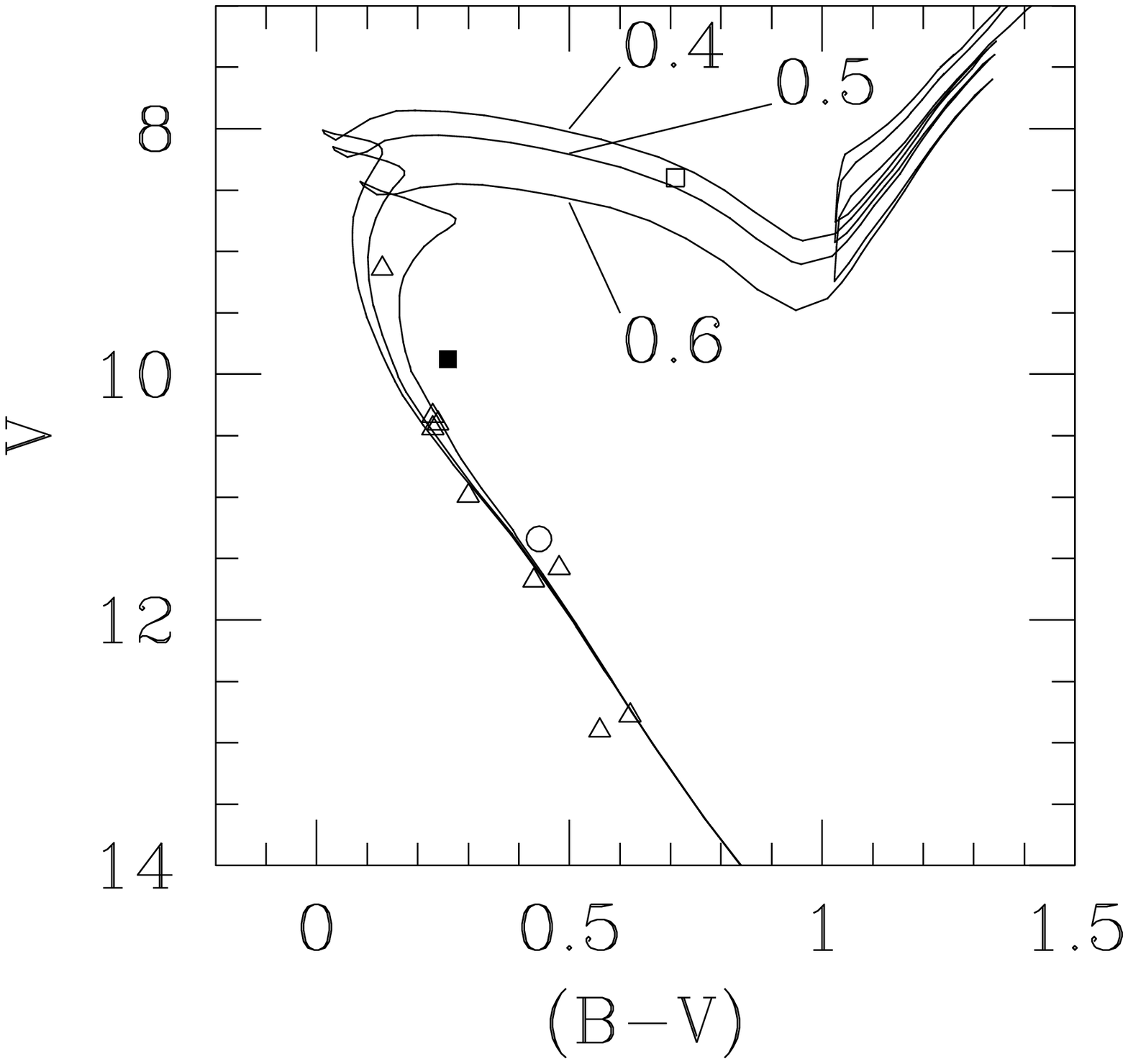}{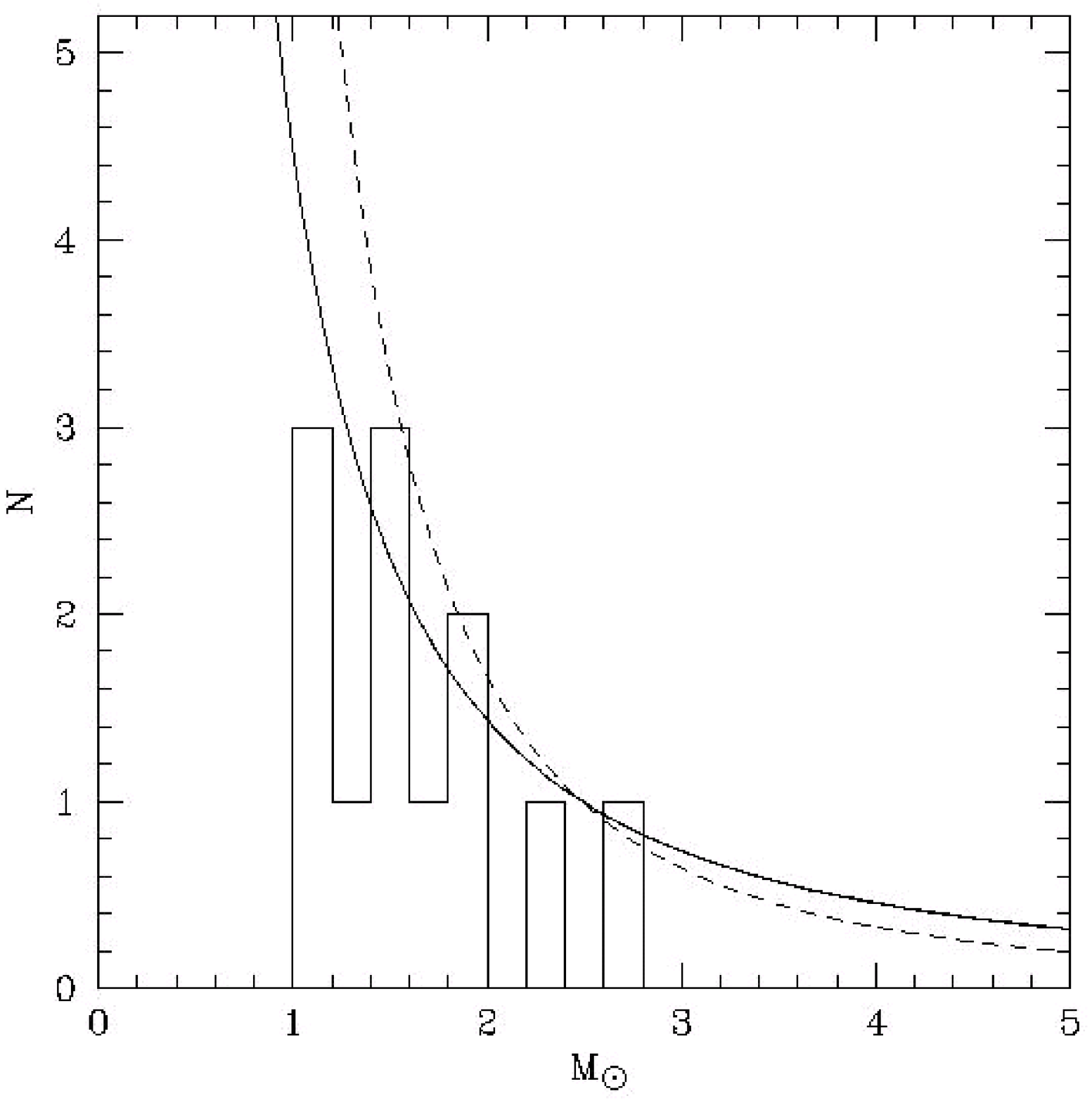}
\caption{{\bf Left}: CMD for cluster members and
age determination. The filled square is a giant
star which actually turns out to be a non member.
{\bf Right}: NGC~1901 Present Day Mass Function.
The dashed line is the classical Salpeter IMF,
whereas the solid line is the best fit to our data.}
\end{figure}

\noindent
Once members have been selected, we build up their
Color Magnitude Diagram (CMD), which is presented in
Fig.~3 (left panel). The cluster exhibits a clear Main
Sequence (MS), which are nicely fit by a solar abundance
500 Myr isochrone from the Padova group (Girardi et al. 2000).
From this fit we derive an absolute distance modulus 
$(m-M)_0$ = 8.25$\pm$0.16
and a reddening E$(B-V)$ = 0.06$\pm$0.03, in agreement with previous
study.
Our spectra are single epoch spectra, and therefore we are not able
to derive a binary fraction estimate, which would be quite an important
constrain for $N$-body models. Second epoch observations are however
planned. Nonetheless, by inspecting the CMD, we can argue that two
stars  off the Main Sequence at V  $\approx$ 11 
might be actually binaries.\\
\noindent
The Present Day Mass Function (PDMF) is not the Salpeter anymore,
but (see Fig.~3, right panel) it is a power law with a slope amounting
at -1.6 (solid line). 
Noticeably, there is a significant lack of low mass
stars with respect to the Salpeter MF (dashed line).\\
The comparison with available $N$-body models of clusters evolution
and dissolution (de la Fuente Marcos 1997) leads us to conclude that
the initial population of the clusters was around 500 stars. In fact
such an initial models after half a billion years should have 15 members,
exactly like NGC~1901.\\

\noindent
Work is in progress to secure membership definition for a number
of POCRs on a radial velocity basis (Villanova 2003, 
Villanova et al. 2003).

\end{document}